%Paper: hep-th/9508178
%From: David M. Pierce <pierce@augustus.physics.unc.edu>
%Date: Thu, 31 Aug 95 16:34:07 EDT
%Date (revised): Fri, 1 Sep 95 11:52:55 EDT

\baselineskip=26pt
\magnification=1200
\vskip .20in
\hskip 1.0in

%\Title{IFP-432-UNC}
{\vbox{\centerline{\bf Threshold Corrections and Gauge}
\vskip2pt\centerline{\bf Symmetry in  Twisted Superstring Models}}}

\smallskip
\centerline{David M. Pierce}
\bigskip\centerline{\it Institute of Field Physics}
\centerline{\it Department of Physics and Astronomy}
\centerline{\it University of North Carolina}
\centerline{\it Chapel Hill, NC 27599-3255, USA}
\bigskip

Threshold corrections to the running of  gauge couplings are calculated for
superstring models with free complex world sheet fermions. For two  N=1
$SU(2)\times U(1)^5$ models, the threshold corrections lead to a small
increase in the unification scale. Examples are given to illustrate how a given
particle spectrum can be described by models with different boundary conditions
on the internal fermions. We also discuss how complex twisted fermions
can enhance the symmetry group of an N=4 $SU(3)\times U(1)\times U(1)$ model to
the gauge group  $SU(3)\times SU(2)\times U(1)$. It is then shown how a mixing
angle analogous to
the Weinberg angle depends on the boundary conditions of the internal fermions.

\bigskip
\bigskip
\centerline{\sl  {\bf 1. Introduction} }
\bigskip

    The unification of gauge coupling constants is a necessary consequence of
string theory. At
tree level, the gauge couplings have simple relations to the string coupling
constant. In higher orders of  perturbation theory, this relation  holds only
at
the Planck mass.  Below this energy, the gauge couplings evolve as determined
by the renormalization group equations. Threshold effects [1] can also modify
the tree level relation and shift the unification scale. Although the effect of
thresholds is small in grand unified field theories, the threshold corrections
can be quite large in string theories [2-7] because there is an infinite tower
of massive states, all of which contribute.

    In this paper, we calculate threshold corrections in four-dimensional
critical superstring models written in terms of free fermions with twisted
boundary conditions.  Complex fermions are useful for studying  theories with
chiral space-time fermions
and for probing the structure of gauge symmetries.  Here, we adapt the
background
field method [5,8] for calculating string thresholds to twisted models in the
framework of type II theories. Although a phenomenologically realistic type II
model has not yet been found, it provides a more economical
construction for using the techniques of low energy string phenomenology.
Calculations of thresholds in heterotic models[5-13] can be made large enough
to
lower the unification scale to an acceptable energy. This  is achieved
by fine tuning the many free moduli parameters of the theory. A desirable
feature of
type II strings is that there is less freedom to adjust the parameters of the
theory.

     In sect. 2, we give a general discussion of the background field method
for running couplings in twisted models, emphasizing models with higher level
Kac-Moody currents. In sect. 3, we calculate the threshold corrections for two
twisted chiral models with $SU(2)\times U(1)^5$ gauge symmetry. These two
models  have the same
massless particle spectrum but different boundary conditions on the internal
fermions. This shows how different boundary conditions on the internal fermions
affect the thresholds. In sect. 4, we investigate the relationship between the
boundary conditions on the internal fermions and the ratio of the field theory
couplings at the Plank mass[14]. Specifically, it will be shown how the twisted
boundary conditions can enhance the symmetry group of an N=4 $SU(3)\times
U(1)\times U(1)$ model to  $SU(3)\times SU(2)\times U(1)$. We then determine a
mixing angle analogous
to the Weinberg angle of the standard model.
\vskip20pt

\centerline{  {\bf 2. Background field calculation}}
\vskip5pt

The tree level relation between gauge couplings is [15]:

$${4\pi\over g_a^2}=2x_a{4\pi\over g_{str}^2}\eqno(2.1)$$
where $x_a$ is the level of Kac-Moody Algebra = N for SU(N). The factor of 2 is
present because we
choose a field theory normalization for the longest roots equal to 1.
This relation is determined by comparing the scattering amplitudes (e.g.
three-point gauge boson vertex)
for the low energy string theory (massless modes) to field theory. The tree
amplitudes are identical if one makes this identification, which holds at  the
Planck mass. We now want to calculate how this relation changes when higher
order corrections and threshold effects are considered. This involves using the
background field method [5,8,17] to find the threshold corrections. Since we
are interested in models with chiral space-time fermions, we consider models
with complex twisted fermions.

     The background field method involves describing the effective action of
the quantum gauge field
as  an effective action of a string propagating in a classical background gauge
field:

$$\Gamma[A_\mu ^a]=\Gamma[X^\mu=0,\Psi ^\mu=0,A_\mu^a]\,.\eqno(2.2)$$

\noindent The effective action will now include contributions from the massive
modes of the string. $X^\mu$,
$\Psi ^\mu= 0$ means that there are no external string states.  In addition,
since the gauge fields are
classical, they do not circulate in loops. In other words, the classical gauge
fields  only exist as external states.
    Polchinski[16] derived a formula for the one-loop correction to (2.1):

$$\Gamma[X^{\mu}=0,\Psi^{\mu}=0,A_{\mu} ^a]=\int d^4x(-{1\over
4g_a^2}F_{\mu\nu}^aF^{a\mu\nu})+\int {d^2\tau\over \tau_2}{\bf Z} +
...\eqno(2.3) $$

\noindent where ${\bf Z}$ is the partition function (one loop with no external
states) of the string in the presence
of a background gauge field. $\tau = \tau_1 + i\tau_2 $ are the coordinates on
the torus.
The first term in (2.3) is simply the classical action of the gauge field and
$-{1\over  4 g_a^2}$ is the tree approximation for the gauge coupling. The
second term, which is the one-loop correction to the tree level result, can
be shown to be equivalent to a one-loop two point string amplitude where the
string
vertex operator for the emission of a gauge boson is modified by making the
substitution $\epsilon_\mu e^{i k\cdot X(z,\bar z)}\rightarrow A_\mu= -{1\over
2}F_{\mu\nu}X^\nu$ provided that $A_\mu$ satisfies the classical equations of
motion. The first order correction to the field theory coupling constants is
then given by the
the coefficient of the $-{1\over 4}F_{\mu\nu}^2 $term in the one-loop two point
background gauge field amplitude.
We now outline this method  for  models containing twisted fermions.

Type II 4-dimensional string models can contain chiral space-time fermions
[18-21] if some of the internal coordinates take values on a shifted lattice
$\sqrt{2\alpha'}p\in Z+\nu$.
These complex twisted fermions satisfy the following boundary conditions:
$$\psi (e^{2\pi i}z)=-e^{2\pi i\nu} \psi (z)\quad ;\quad  \tilde \psi (e^{2\pi
i}z) =-e^{-2\pi i\nu}\tilde \psi (z)\eqno(2.4)$$
where $\nu$ is real. The space-time fermions are real, either Neveu-Schwarz or
Ramond. They are given by:
$$\psi^\mu(z)=\sum_{r\in Z+{1\over 2}}\psi_r^\mu z^{-r-{1\over 2}}\quad ;
\qquad\psi^\mu(z)=\sum_{n\in Z}\psi_n^\mu z^{-n-{1\over 2}}\eqno(2.5)$$
for Neveu-Schwarz and Ramond respectively.
Twisted models differ from untwisted models in that internal fermions can be
defined by the following for any value of $\lambda$
 $$\psi^i(z)=\sum_{r\in Z+\lambda}\psi_r^i z^{-r-{1\over 2}}\quad ;\quad \tilde
\psi^i(z)=\sum_{r\in Z-\lambda}\tilde \psi_r^i z^{-r-{1\over
2}}\eqno(2.6)$$where $\lambda = {1\over 2} -\nu$ and $f_r^\dagger= \tilde
f_{-r}$. $\nu = 1/2$ is the Ramond case and $\nu= 0$ the Neveu-Schwarz case.
\vskip1pt
     We consider 4-dimensional type II models in the light-cone description.
Here, the left and right movers can each be described by two bosonic and twenty
fermionic fields. The partition function without the zero modes p is given by
$$\eqalign{Z&=\prod_{l=0}^{k-1}{1\over {\cal
N}_l}\sum_{\alpha,\beta}c(\alpha,\beta)
Tr_\alpha[q^{L_0'-{1\over 2}}\bar q^{\bar L_0'-{1\over
2}}(e^{-i\pi})^{\rho_\beta\cdot F}]\cr &=
\prod_{l}{1\over {\cal N}_l}
\sum_{\alpha,\beta}c(\alpha,\beta)
|\eta(q)|^{-24}
\prod_{i=1}^{n}
\left(\vartheta\left[{\rho_\alpha^i\atop\rho_\beta^i}\right](0|q)\right)^{1/2}
\prod_{i=1}^{n'}\left(\vartheta\left[{\bar\rho_\alpha^i\atop\bar\rho_\beta^i}
\right](0|\bar q)\right)^{1/2}\cr
&\times\prod_{i=1}^{m}\vartheta\left[{\rho_\alpha^i\atop\rho_\beta^i}\right](0|q)\prod_{i=1}^{m'}\bar \vartheta\left[{\rho_\alpha^i\atop\rho_\beta^i}\right](0|\bar q)\cr}\eqno(2.7)$$
where the prime on the Hamiltonian denotes the omission of the bosonic zero
modes. In this formalism, the partition function is
 a sum over the sectors generated by $\rho_\alpha\equiv(\rho_\alpha;\bar
\rho_\alpha)\in\Omega$. Each sector $\alpha$ contains $n+n'$ real fermions and
$m+m'$ complex fermions. $\rho_\alpha$ is a ($n+n'+m+m'$) dimensional vector
which describes the boundary conditions of the fermions for each sector:
$$\rho_\alpha
=2(\nu_1,...\nu_n;\nu_1,...\nu_m;\nu_1,...\nu_{n'};\nu_1,...\nu_{m'})\,.\eqno(2.8)$$
$c(\alpha,\beta)=\delta_\alpha\epsilon(\alpha,\beta)$ are phases for the
$(e^{-i\pi})^{\rho_\beta\cdot F}$ projections.
F is a vector whose components are the operators $F_j=\sum_{r\in
z+\lambda}:f_r^j\tilde f_{-r}^j:$ for complex fermions and $\sum_{s=1/2}^\infty
b_{-s}^j b_s^j$
or $\sum_1^\infty d_{-n}^j d_n^j$ for real NS or R fermions respectively.
$$\rho_\beta\cdot F =2\sum_{j=1}^n\nu_jF_j^L + 2\sum_{j=1}^m\nu_jF_j^L
-2\sum_{j=1}^{n\prime} \nu_jF_j^R -2\sum_{j=1}^{m\prime} \nu_j
F_j^R\,.\eqno(2.9)$$ In addition, the factor $\prod_{l=0}^{k-1} {\cal N}_l$ (
k= the number of generators) is  the number of sectors where the order, ${\cal
N}_\alpha$ is defined by $\alpha^{{\cal
N}_\alpha}=\phi=((1)^n;(1)^m;(1)^{n'};(1)^{m'})$
and $\alpha$ is a  vector whose components are given by
$$\alpha =(e^{2\pi i\nu},...;e^{2\pi i\nu},...;e^{2\pi i\nu},...;e^{2\pi
i\nu},...)\eqno(2.10)$$

The generalized Jacobi theta functions are given by

$$\vartheta\left[{\rho\atop\mu}\right](\nu|\tau)=\sum_{n\in
Z}e^{i\pi\tau(n+\rho/2)^2}e^{-i2\pi(n+\rho/2)(\nu+\mu/2)}e^{i\pi\rho\mu/2}$$

$$\bar \vartheta\left[{\bar \rho\atop\bar \mu}\right](\nu | \bar
\tau)=\sum_{n\in Z}e^{-i\pi\bar \tau(n+\bar \rho/2)^2}e^{i2\pi(n+\bar
\rho/2)(\nu+\bar \mu/2)}e^{-i\pi\bar \rho \bar \mu/2}\eqno(2.11)$$
with $q=e^{2\pi i \tau}$, $\bar q =e^{-2\pi i \bar \tau}$,$\tau= \tau_1
+i\tau_2 $, and $\bar \tau =\tau_1-i\tau_2$.
Note that this differs from that given in [18] because the left movers are now
functions of q rather that $\bar q $.

 The one-loop two point amplitude contribution to the effective lagrangian for
$A_\mu^a$
background is :
$${\cal L}'(A_\mu^a)=\prod_l {{1\over\cal N}_l} \sum_{\alpha,\beta}
c(\alpha,\beta) \int {d^4p\over
(2\pi)^4}Tr_\alpha[\Delta V^a(1,1)\Delta V^a(1,1)(e^{-i\pi})^{\rho_\beta\cdot
F}]\eqno(2.12)$$
where the sum over sectors corresponds to a generalized GSO
projection[18]. The closed string propagator is

$$\Delta={1\over4\pi}\int_{|z|\le1}{dz d\bar z\over|z|^2}
z^{L_0-{1\over 2}}\bar z^{\bar L_0-{1\over 2}}\,.\eqno(2.13)$$
The Hamiltonian $L_0$ and associated Virasoro generators are given by:
$$L_n=\sum_{r\in Z+\lambda}(r-{n\over 2}):\tilde f_{n-r}f_r: +{1\over
4}\sum(\lambda-{1\over 2})^2 \delta_{n.0}\,.\eqno(2.14)$$
Recall that the background field method involves making the substitution
$\epsilon_\mu e^{ik\cdot X(z,\bar z)}\rightarrow A_\mu(x)$ in the vertex
operator for a gauge boson
provided that $A_\mu(x)$ satisfies the equation of motion $\partial_\mu
F^{\mu\nu}=0.$
The vertex operator for a gauge boson, $b^{La}_{-{1\over 2}} \epsilon\cdot
b^R_{-{1\over 2}}|0>$, is constructed in part from the Kac-Moody currents of
(2.15) and (2.16).  In models with complex fermions, such as the two chiral
models
that are discussed in sect. 3, the affine algebra is constructed for a
particular gauge group. For a model with complex fermions and an $SU(2)\times
U(1)^5$
gauge symmetry, the currents  are given by:
$$J^a(z)=\sum_{n\in Z}J_n^a
z^{-z}=-{i\over2}f_{abc}\psi^b(z)\psi^c(z)\eqno(2.15)$$
where $f_{abc}$ are the structure constants of $SU(2)$ and $3\le a,b,c \le 5$.
\vskip1pt
\noindent The U(1) currents are:
$$ J^a(z)=  :f^j(z)\tilde f^j(z): + \nu_j\eqno(2.16)$$
for $6\le a\le 10$, $1\le j\le 5$, and no sum on j.
\noindent The zero modes of these  currents  generate  the
gauge symmetry. The vertex operator
for an $SU(2)$ gauge boson is:
$$\eqalign{V^a(k,\epsilon,z,\bar z)=&[{\textstyle {1\over2}}k\cdot
\psi^L(z)\psi^{La}(z)- {\textstyle{{i\over 2}}}f_{abc} \psi^{Lb}(z)
 \psi^{Lc}(z)]\cr
&\epsilon\cdot[i\bar z\bar \partial X_{\rm R}(\bar z)-
{\textstyle {1\over2}}\psi^R(\bar z) k\cdot\psi^R(\bar z)]e^{ik\cdot
X(z,\bar z)}\cr}\eqno(2.17)$$
Here, all the fermionic oscillators are real.
On the other hand, the vertex operators for the U(1) gauge bosons are
constructed from the corresponding Kac-Moody currents :
$$\eqalign{V^a(k,\epsilon,z,\bar z)=&[{\textstyle {1\over2}}k\cdot
\psi^L(z)\psi^{L a}(z)+ : \psi^{L j}(z)
 \tilde \psi^{L j}(z):]\cr
&\epsilon\cdot[i\bar z\bar \partial X_{\rm R}(\bar z)-
{\textstyle{ 1\over2}}\psi^R(\bar z) k\cdot\psi^R(\bar z)]e^{ik\cdot
X(z,\bar z)}\cr}\eqno(2.18)$$
where  $\nu_j$ in the current from (2.16) is zero.
The vertex operator also contains the bosonic fields given by
$$\eqalign{X^\mu(z,\bar z)&={x^\mu}
+{p^\mu\over4i}(\ln z+\ln\bar z)
+{i\over2}\sum_{n\ne0} {1\over n}\tilde\alpha_n^\mu z^{-n}
+{i\over2}\sum_{n\ne0} {1\over n}\alpha_n^\mu \bar z^{-n}\cr
&={1\over2}\left(X_{\rm L}^\mu(z)+X_{\rm R}^\mu(\bar z)\right)\cr}\eqno(2.19)$$
 It is only necessary to evaluate  one component of a non-abelian subgroup at a
time. For example, for $SU(2)$, one would look at only one of the three gauge
bosons . Due to gauge invariance, it doesn't matter which one is selected.
For a constant $F_{\mu\nu}$ corresponding to a given
component of subgroup a, the resulting background field vertex is

$$\eqalign{V^a[F_{\mu\nu}](z,\bar z)={\textstyle{i\over4}}F_{\mu\nu}&
\bigl\{
J^a(z)[2X^\mu(z,\bar z)\bar z\bar\partial X_{\rm R}^\nu(\bar
z)-\psi^{R\mu}(\bar z)
\psi^{R\nu}(\bar z)]\cr
&-i[\psi^{L\mu}(z) \psi^{La}(z)][\bar z
\bar\partial X_{\rm R}^\nu(\bar z)]\bigr\}\cr}\eqno(2.20)$$
where the gauge currents are given in  (2.15) and (2.16).
Using operator methods one can rewrite (2.12) as:
$$\eqalign{{\cal L}'(A_\mu^a)&=\prod_l {1\over{\cal N}_l} \sum_{\alpha,\beta}
c(\alpha,\beta)\pi^2 \int {d^4p\over
(2\pi)^4}\int_\Gamma d^2\tau \int_ {0\leq Im\nu \leq Im\tau}d^2\nu \cr &\times
Tr_\alpha[ V^a(z,\bar z)V^a(1,1)q^{L_0-1/2}\bar q ^{\bar L
_0-1/2}(e^{-i\pi})^{\rho_\beta\cdot F}]}\eqno(2.21)$$
where z=$e^{2\pi i\nu}$, $q=e^{2\pi i\tau}$, and the integration is restricted
to the fundamental region $\Gamma$. Performing the trace:
$$\eqalign{Tr&_\alpha [V^a(z,\bar z)V^a(1,1)q^{L_0-1/2}
\bar q ^{\bar L_0-1/2}(e^{-i\pi})^{\rho_\beta \cdot F}]=\cr &
-{\textstyle{1\over16}}F_{\mu\nu}F_{\rho\sigma}Tr_\alpha [q^{L_0-1/2}\bar
q^{\bar L_0-1/2}(e^{-i\pi})^{\rho_\beta \cdot F}]\cr & \times \{\langle
J^a(z)J^a(1)\rangle  [\langle 2X^\mu(z,\bar z)\bar z \bar \partial
 X^{R\nu}(\bar z)2X^\rho(1,1)\bar \partial X^{R\sigma}(1)\rangle
 + \langle \psi^\mu(\bar z)\psi^\nu(\bar z)\psi^\rho(1)\psi^\sigma (1)\rangle
]\cr & -\langle \psi ^{L\mu}(z)\psi ^{La}(z)\psi^{L\rho}(1)\psi^{La}(1)\rangle
\langle \bar z \bar \partial X^{R\nu}(\bar z)\bar \partial
X^{R\sigma}(1)\rangle \}\cr }\eqno(2.22)$$
where the two point correlation function on a torus is defined by:
$$\langle A(z,\bar z)B(w,\bar w)\rangle \equiv {{Tr_\alpha[A(z,\bar z)B(w,\bar
w)q^{L_0-1/2}\bar q^{\bar L_0-1/2}(e^{-i\pi})^{\rho_\beta \cdot F}
]}\over{Tr_\alpha[q^{L_0-1/2}\bar q^{\bar L_0-1/2}(e^{-i\pi})^{\rho_\beta \cdot
F} ]}}\eqno(2.23)$$
The last term is a total derivative and  drops out after using
the generalized Gauss's theorem. After performing the p integration
we have
$$\eqalign{{\cal L}'(&A_\mu^a)={\textstyle{1\over4}}F_{\mu\nu}^2{1\over
16\pi^2}\prod_l {1\over {\cal N}_l} \sum_{\alpha,\beta}
c(\alpha,\beta)\int_\Gamma {d^2\tau\over \tau_2}\int {d^2\nu\over
\tau_2}\langle J^a(z)J^a(1)\rangle _{\alpha,\beta} \cr &\times 2[\langle
\Psi(\bar z)\Psi (1)\rangle ^2_{\alpha,\beta}-\langle X^R(
\bar z)\bar z\bar \partial X^R(1)\rangle
^2_{\alpha,\beta}]Tr_\alpha[q^{L'_0-1/2}\bar q ^{\bar L'
_0-1/2}(e^{-i\pi})^{\rho_\beta\cdot F}]\cr}\,.\eqno(2.24)$$
The fermionic current gives rise to a gauge independent(apart from $k_i$) part
and a gauge dependent part:
$$\langle J^a(z)J^a(1)\rangle_{\alpha,\beta}= -k_a(z{\partial\over {\partial
z}})^2\log\theta_1(z,q) + \langle J_0^aJ_0^a\rangle\,.\eqno(2.25)$$
Since $k_a$ is one, the first part will shift all the groups by the same amount
and can thus be absorbed into a redefinition of the string coupling constant.
Using the explicit expressions for the  currents, their correlation functions
are found to be:
$$\langle J^a_0J^a_0\rangle_{\alpha,\beta}= {1\over
2}f_{cd}^af_{cd}^a2q\log\theta\left[{\rho_\alpha^c \atop \rho_\beta
^c}\right](0|q)\eqno(2.26)$$
for the SU(2) currents(no sum on a) and
$$\langle J^a_0J^a_0\rangle_{\alpha,\beta}=2q\log\theta\left[{\rho_\alpha^j
\atop \rho_\beta ^j}\right](0|q)\eqno(2.27)$$
for the U(1) currents($1\le j\le 5$).
\vskip1pt
\noindent In the $SU(3)\times U(1)\times U(1)$ model  presented in sect. 4, the
currents
are given by:
$$J^a(z)=\bar J^a(z)+q^a(z)$$
$$\bar J^a(z)=-{i\over 2}f_{abc}b^b(z)b^c(z)\quad  ;\quad
q^q(z)=i\lambda^a_{ij}:f^i(z)\tilde f^j(z):\eqno(2.28)$$
$$J^{11}(z)={1\over \sqrt{3}}[ :f^i(z)\tilde f^i(z) : + 3\nu_2]\quad ;\quad
J^{12}(z)=[:f^1(z)\tilde f^1(z) : + \nu_1]$$
where $3\le a,b,c \le10$, $2\le i,j\le 4$, and the strucure constants of SU(3)
are normalized so $C_{\psi}=2$.
\vskip1pt
\noindent The correlation function of the $SU(3)$ current is(no sum on a):

$$\langle J_0^aJ_0^a\rangle_{\alpha,\beta}=  {1\over
2}f_{cd}^af_{cd}^a2q\log\theta\left[{\rho_\alpha^c \atop \rho_\beta
^c}\right](0|q)+\lambda_{ij}^a{\lambda_{ij}^a}^*2q\log\theta\left[{\rho_\alpha^j \atop \rho_\beta ^j}\right](0|q)\,.\eqno(2.29)$$

Performing the $\nu$ integration on the gauge dependent part,we have:
$$\eqalign{{\cal L}'(A_\mu^a)&={\textstyle{1\over 4}}F_{\mu\nu}^2 {1\over
16\pi^2}\prod_{l}{1\over {\cal N}_l}
\sum_{\alpha,\beta}c(\alpha,\beta)\int_\Gamma{d^2\tau\over\tau_2}2\langle
J_0^aJ_0^a\rangle\cr  &\times 2\bar q{d\over d\bar q}
\log\left(
{{\left (\bar \vartheta\left[\bar\rho^1_\alpha\atop\bar\rho^1_\beta\right](\bar
q)\right )^{1\over2}\left (\bar
\vartheta\left[\bar\rho^2_\alpha\atop\bar\rho^2_\beta\right](\bar q)\right
)^{1\over 2}}
 \over
\eta(\bar q)}
\right)
Tr_\alpha[q^{L_0'-{1\over 2}}\bar q^{\bar L_0'-{1\over 2}}
(e^{-i\pi})^{\rho_\beta\cdot F }]\cr}\eqno(2.30)$$
Now using the expression for the partition function for
twisted fermions (2.4),
we see that the one-loop two point background gauge field contribution to the
effective lagrangian is
$${\cal
L}'(A_\mu^a)=-{\textstyle{1\over4}}F_{\mu\nu}^2{1\over16\pi^2}\int_\Gamma
{d^2\tau\over\tau_2}[2B_a(q,\bar q)+Y']\eqno(2.31)$$
where $Y'$ is the gauge independent part and
$$\hskip-5pt\eqalign{B_a(q,\bar q)&=-\prod_{l}{1\over {\cal
N}_l}\sum_{\alpha,\beta}
c(\alpha,\beta)
{|\eta(q)|^{-22}\over\eta(q)}
2\bar q{d\over d\bar q}\left(\left (\bar
\vartheta\left[{\bar\rho^1_\alpha\atop\bar\rho^1_\beta}
\right](\bar q) \bar \vartheta\left[{\bar\rho^2_\alpha\atop\bar\rho^2_\beta}
\right](\bar q)\right )^{1\over2}
\bigg/\eta(\bar q)\right)\langle J_0^aJ_0^a\rangle\cr &\times
\prod_{j=1}^{n}\left(\vartheta\left[{\rho^j_\alpha\atop\rho^j_\beta}\right](q)\right)^{1/2}\prod_{j=3}^{n\prime}\left(\bar \vartheta
\left[{\bar\rho^j_\alpha\atop\bar\rho^j_\beta}\right](\bar
q)\right)^{1/2}\prod_{j=1}^{m}\vartheta\left[{\rho^j_\alpha\atop\rho^j_\beta}\right](q)\prod_{j=1}^{m\prime}\bar \vartheta\left[{\bar\rho^j_\alpha\atop\bar\rho^j_\beta}\right](\bar q)\cr}\eqno(2.32)$$

The first order correction to the field theory coupling is given by the
coefficient of $-{1\over4}F_{\mu\nu}^2$ in the one-loop two point background
gauge field amplitude (2.31).
In this analysis, we use the type II string normalization of $C_{\psi}=2$. To
compare this result with field theory, which uses a normalization of
$C_{\psi}=1$, we must multiply this result by a factor
$\Psi^2_{FT}/\Psi^2_{str}=x_a/2$. Here, $\Psi^2$ is the length squared of the
longest root in the subgroup and $x_a$ is the level of the Kac-Moody algebra.
Then, following ref.[5], we get an equation for the  $\overline{DR}$ couplings
$${1\over\alpha_i(\mu)}={x_a/2\over\alpha_{\rm
GUT}}-{b_a\over{2\pi}}\log\mu/M_{\rm str}+{\Delta_a\over4\pi}\,.\eqno(2.33)$$
The gauge independent part has been absorbed into a redefinition of the
string coupling by
$${1\over\alpha_{\rm GUT}}={4\pi\over g_{str}^2}+{Y\over{4\pi}}\eqno(2.34)$$
where $Y=\int_\Gamma{d^2\tau\over\tau_2}Y'$ and $\alpha_a=g^2_a/4\pi$.
The massive string contributions are given by the thresholds $\Delta_a$:
$$\Delta_a=\int_\Gamma{d^2\tau\over\tau_2}[x_aB_a(q,\bar
q)-b_a]\,.\eqno(2.35)$$
The $b_a$ are the field theory $\beta$ functions given by[22]
$$b_a=-{11\over 3}Tr_{\rm V}(Q_i^2)
+{2\over 3}Tr_{\rm F}(Q_i^2)+{1\over 6}Tr_{\rm S}(Q_i^2)\eqno(2.36)$$
Here, the traces are over two-component fermions and real scalars(in field
theory normalization). The massless contribution from the string is given by
$$b_a=\lim_{q\to 0}x_aB_a(q,\bar q)\eqno(2.37)$$
and should equal the field theory result. This provides a consistency check for
the string calculation.
One can facilitate the numerical integration by making a change of variables:
$\tau_2\equiv Im\tau=1/\tau_2'$. Since
$B_a(-\tau_1,\tau_2)=B(\tau_1,\tau_2)^\ast$, the imaginary part drops out
leaving:
$$\Delta_a=-2Re\int_0^{.5}d\tau_1\int_{1/\sqrt{1-\tau_1^2}}^0{d\tau_2'\over\tau_2'}[x_aB_a(\tau_1,\tau_2')-b_a]\,.\eqno(2.38)$$
\vskip8pt

\centerline{{\bf 3. Threshold calculation for two  N=1 $SU(2)\times U(1)^5$
chiral models}}
\vskip5pt
We now calculate the threshold corrections for two twisted models with
$SU(2)\times U(1)^5$ symmetry. In both cases, the thresholds increase the
unification scale  by a very small amount.
\vskip1pt
\noindent {\bf Example 1}

\noindent
A model with $SU(2)\times U(1)^5$ gauge symmetry [23,19] can be described by
three generators $b_0,b_1,b_2: {\cal N}_0={\cal N}_1=2: {\cal N}_2$=4: K=2. The
sixteen sectors of the model  can be determined from the vectors $\rho_{b_i}$
describing the generators:

         $$\eqalign{\rho_{b_0}&=((1)^{12};(1)^4;(1)^{12};(1)^4)\cr
  \rho_{b_1}&=((0)^{12};(0)^4;(1)^4(0)^8;(0)^2(1)^2)\cr
\rho_{b_2}&=((0)^{10}(1)^2;(1/2)^4;(0)^2(1)^2(0)^4(1)^4;(1/2)^4)\cr}\eqno(3.1)$$
\vskip1pt
\noindent Figure 1 illustrates the boundary conditions for the sectors of the
model. The massless states satisfy the following criteria:
\vskip2pt
\noindent (1): For the states to survive the projections, we must have:

$$ e^{-i\pi \rho_{b_i}\cdot F}\alpha=\epsilon
(\alpha,b_i)^\ast\alpha\eqno(3.2)$$
(2): The left and right movers must each have a mass eigenvalue of 0:
$$ \alpha' {m_L}^2|S>=(L^L_0-1/2)|S>=0\quad ;\quad \alpha'
{m_R}^2|S>=(L^R_0-1/2)|S>=0\eqno(3.3)$$

\noindent The massless states come from the sectors: $b_1$,
$b_2$,mm
$b_1b_2$,
$b_2^2$, $b_1b_2^2$, $b_2^3$, $b_1b_2^3$.
For $g = SU(2)\times U(1)^5$, they are:
\vskip6pt
\line{\hskip 3mm 1) from the untwisted sector,\hfil}
\vskip3pt
\settabs\+\hskip 6mm&spin &$(\pm {1\over 2},2(0))$\quad & in \cr
\+&spin &$(\pm2,\pm{3\over2})$ &in (1;0,0,0,0,0) of
$g$\cr
\vskip1pt
\+&&$(\pm1,\pm{1\over2})$ &in adjoint of $g$\cr
\+&&$({1\over2},0)$ &in $(1;\pm1,0,0,0,0)\oplus
2(1;0,\underline{1,0,0,0})$ of $g$\cr
\+&&$(-{1\over2},0)$ &in $(1;\pm1,0,0,0,0)\oplus
2(1;0,\underline{-1,0,0,0})$ of $g$ \cr
\+&&$(\pm{1\over2}, 2(0))$ &in $(1;0,0,0,0,0)$ of $g$\cr
\vskip6pt
\line{\hskip 3mm 2) from the singly twisted sector,\hfil}
\vskip 3pt
\+&&$({1\over 2},0)$ &in $2(1;\pm{1\over2},
\underline{-
{3\over4},{1\over4},{1\over4},{1\over4}})$ of $g$\cr
\vskip6pt
\line{\hskip3mm 3) from the anti-twisted sector,\hfil}
\vskip3pt
\+&&$(-{1\over2},0)$ &in $2(1;\pm{1\over2},
\underline{{3\over4},-{1\over4},-{1\over4},-
{1\over4}})$ of $g$\cr
\vskip6pt
\line{\hskip 3mm 4) from the doubly twisted sector,\hfil}
\vskip3pt
\+&&$({1\over2},0)$ &in $(1;0,
\underline{{1\over2},{1\over2},-{1\over2},-
{1\over2}})\oplus
(1;0,{1\over2},{1\over2},{1\over2},{1\over2})$\cr
\+&&&\hskip 5mm $\oplus (1;0,-{1\over2},-{1\over2},-{1\over2},-
{1\over2})
\oplus 2(1;0,\underline{
{1\over2},-{1\over2},-{1\over2},-{1\over2}})$ of $g$\cr
\+&&$(-{1\over2},0)$ &in $(1;0,\underline{-{1\over2},
-{1\over2},{1\over2},{1\over2}})\oplus
(1;0,-{1\over2},-{1\over2},-{1\over2},-{1\over2})$\cr
\+&&&\hskip5mm $\oplus (1;0,{1\over2},{1\over2},{1\over2},{1\over2})
\oplus 2(1;0,\underline{-{1\over2},{1\over2},
{1\over2},{1\over2}})$ of
$g$.\cr
\vskip 2pt

\vskip20pt
\noindent {\bf Example 2}
\vskip1pt
\noindent Example 2 describes an N=1 $SU(2)\times U(1)^5$ model. The
generators  are :

$\rho_{b_0}=((1)^{10};(1)^2(1)^2(1)^6;(1)^4(1)^4;(1)^4(1)^4)$
\vskip1pt

$\rho_{b_1}=((0)^{10};(0)^2(0)^2(0)^6;(1)^2(1)^2(0)^4(0)^4;(0)^4(0)^4)\hskip100pt (3.4)$
\vskip1pt

$\rho_{b_2}=((0)^{10};(3/4)^2(1/2)^2(1/4)^6;(0)^2(1)^2(0)^4(1)^4;(1/2)^4(1/2)^4)$
\vskip1pt
\noindent Figure 2 represents the boundary conditions on the fermions in each
sector. The massless spectrum is :
\vskip2pt
Sectors $b_1,\phi$:
\vskip2pt
spin $(\pm 2,\pm {3\over 2})$ \hskip5pt in  (1;0,0,0,0,0)  of g
\vskip1pt
\hskip15pt    $(\pm 1,\pm {1\over 2})$\hskip5pt in  adjoint of g
\vskip1pt
\hskip15pt    $(\pm {1\over 2},2(0))$\hskip5pt  in  (1;0,0,0,0,0) of g
\vskip1pt
\hskip15pt     $(0,{1\over 2})$ \hskip5pt  in 2(1;1,0,0,0,0)
\vskip1pt
\hskip15pt     $(0,-{1/2})$ \hskip5pt  in 2(1;-1,0,0,0,0)
\vskip2pt
Sectors $b_2,b_2b_1,b_2^7,b_2^7b_1$
\vskip2pt
spin $(0,{1\over 2})$\hskip10pt in \hskip5pt$2(1;-{3\over 4},-{5\over
8},{1\over 8},{1\over 8},{1\over 8})\oplus 2(1;{1\over 4},{3\over
8}\underline{-{7\over 8},{1\over 8},{1\over 8}})$
\vskip 1pt
\hskip15pt
 $(0,-{1\over 2})$\hskip10pt in \hskip5pt $2(1;{3\over 4},{5\over 8},-{1\over
8},-{1\over 8},-{1\over 8})\oplus 2(1;-{1\over 4},-{3\over
8},\underline{{7\over 8},-{1\over 8},-{1\over 8}})$
\vskip 2pt
Sectors  $b_2^3,b_2^3b_1,b_2^5,b_2^5b_1$
\vskip 1pt

spin  $(0,{1\over 2})$\hskip10pt  in \hskip5pt$2(1;-{3\over 4},-{1\over
8},-{3\over 8},-{3\over 8},-{3\over 8})\oplus 2(1:{1\over4},-{1\over
8},\underline{{5\over8},{5\over8},-{3\over8}})$
\vskip 1pt
\hskip20pt
      $(0,-{1\over 2})$\hskip10pt  in\hskip5pt$ 2(1;{3\over 4},{1\over
8},{3\over 8},{3\over 8},{3\over 8})\oplus 2(1;-{1\over 4},{1\over
8},\underline{-{5\over 8},-{5\over 8},{3\over 8}})$
\vskip 1pt
Sectors  $b_2^4,b_2^4b_1$
\vskip 1pt
spin  $(2(0),\pm {1\over 2})$  in $(1;0,{1\over 2},-{1\over 2},-{1\over
2},-{1\over 2})\oplus(1;0,-{1\over 2},{1\over 2},{1\over 2},{1\over 2})$
\vskip 1pt
\hskip20pt
      $(0,{1\over 2})$\hskip10pt  in\hskip5pt   $2(1;0,-{1\over
2},\underline{{1\over 2},-{1\over 2},-{1\over 2}})$
\vskip 1pt

\hskip20pt     $(0,-{1\over 2})$\hskip10pt in \hskip5pt
$2(1;0,{1\over 2},\underline{-{1\over 2},{1\over 2},{1\over 2}})$
\vskip 1pt

Sectors $b_2^2,b_2^2b_1,b_2^6,b_2^6b_1$

spin  $(0,{1\over 2})$\hskip10pt    in\hskip5pt   $(1;-{1\over 2},{3\over
4},{1\over 4},{1\over 4},{1\over 4})\oplus 2(1;{1\over 2},{3\over 4},{1\over
4},{1\over 4},{1\over 4})$
\vskip 1pt
\hskip100pt
                     $\oplus(1;-{1\over 2},-{1\over 4},\underline{-{3\over
4},{1\over 4},{1\over 4}})$
\vskip1pt
\hskip20pt
 $(0,-{1\over 2})$\hskip10pt in \hskip5pt   $(1;-{1\over 2},{3\over 4},{1\over
4},{1\over 4},{1\over 4})\oplus 2(1;{1\over 2},{1\over 4},\underline{-{3\over
4},{1\over 4},{1\over 4}})$
\vskip1pt
\hskip100pt
$\oplus(1;-{1\over 2},-{1\over 4},\underline{-{3\over 4},{1\over 4},{1\over
4}})$
\vskip1pt
\hskip20pt $(0,{1\over 2})$\hskip10pt  in\hskip5pt    $(1;{1\over 2},-{3\over
4},-{1\over 4},-{1\over 4},-{1\over 4})\oplus 2(1;-{1\over 2},{1\over
4},\underline{{3\over 4},-{1\over 4},-{1\over 4}})$
\vskip 1pt
\hskip100pt    $\oplus(1,{1\over 2},{1\over 4},\underline{{3\over 4},-{1\over
4},-{1\over 4}})$

\hskip20pt  $(0,-{1\over 2})$\hskip10pt in \hskip5pt       $(1;{1\over
2},-{3\over 4},-{1\over 4},-{1\over 4},-{1\over4})\oplus2(1;-{1\over
2},-{3\over 4},-{1\over 4},-{1\over 4},-{1\over 4})$
\vskip 1pt
\hskip100pt    $\oplus(1;{1\over 2},{1\over 4},\underline{{3\over 4},-{1\over
4},-{1\over 4}})$

\bigskip

\noindent{\bf Equivalence of the two $SU(2)\times U(1)^5$ models}

The quantum numbers of each state
are  described by a six component vector in an $SU(2)\times U(1)^5$ space.
The equivalence of the massless spectra is most easily shown by determining the
length and relative orientation of the vectors in their respective spaces.
Since all the states are $SU(2)$ singlets, we only need to look at the $U(1)$
quantum numbers. Comparing the two $U(1)^5$ spaces, one finds that
every non-zero vector has a magnitude of one. There are also an equal number of
pairs of vectors having the same angular separation in both
spaces. Thus we see  that they are related by a rotation of  the coordinates
which describe the
quantum numbers of the massless states. To be more specific, the rotations that
take the
first model into the second model are described by a 5x5 transformation matrix.
If $S_1$ is
a state in the first model :
$$S_1=\left(\matrix{
U_1\cr
U_2\cr
U_3\cr
U_4\cr
U_5\cr
}\right)\eqno(3.5)$$ a state in the second model $S_2$ is given by:
$$S_2=M S_1\eqno(3.6)$$
where M  is the transformation matrix given by:
$$M={1\over 2}\left(\matrix{
\hskip6pt 0  &  2  & \hskip6pt 0  & \hskip6pt 0  & \hskip6pt  0 \cr-1 &  0  &
-1  &  -1 &  -1 \cr
\hskip6pt1 &  0 & -1  &\hskip6pt  1 & -1 \cr
\hskip6pt 1 &  0 & -1  & -1  & \hskip6pt 1 \cr
\hskip6pt 1 &  0 & \hskip6pt 1 & -1  & -1 \cr
}\right)\,.\eqno(3.7)$$

To compare the massive spectra, we calculate the partition function for
both models.  Although the partition function can be shown to be
identically zero for both models, one can look at the number of massive states
at each level for bosons and fermions separately.  For supersymmetric theories,
these contributions cancel. When we expand the partition function we obtain for
the bosons of both models:
$$Z=104+24\bar q^{1\over 2}q^{-{1\over 2}}+256\bar q^{-{1\over 4}}q^{3\over 4}
+ 1536 q^{1\over 4}\bar q^{1\over4} +... \eqno(3.8)$$
\noindent Both models have the same massless spectrum and appear to have the
same
number of massive states. This suggests that they are  equivalent.

\bigskip
\noindent{\bf Thresholds}
\vskip1pt

 For example 1, the levels of the
gauge groups are given by $x_a=$ (2;2;2,2,2,2). Using these values, we find
that the zero-mode contribution of the string (2.37) agrees with the field
theory  (2.36). These $\beta$-functions  are given by $b_a=$ (-6;6;9,9,9,9)
for the five gauge groups.  The presence of $b_a$
ensures that the integral in (2.38) converges.
 A mathematica computer program was developed to calculate the expression
(2.38) and to perform the numerical integrations. Each sector was evaluated
independently to check for  consistency with field theory.  As a result we find
that the thresholds for example 1 are $\Delta_a=$ (-1.25; 0.074, 0.41, 0.41,
0.41, 0.41).

   We would now like to know how the massive modes of the string will change
the unification scale.  For this particular chiral model, it appears
that there is a small increase in the unification scale. If we include the
massive string contributions, the equations describing the running of the field
theory coupling constants are given by:
$$ {1\over \alpha_2 (\mu)} ={1\over\alpha_G}  + {3\over\pi}\log({\mu \over
M_{st}}) -{1.25\over 4\pi}$$
$$ {1\over \alpha_{1a} (\mu)}={1\over\alpha_G} -{3\over\pi}\log({\mu \over
M_{st}}) + {.074\over 4\pi}\eqno(3.9)$$
$$ {1\over \alpha_{1b} (\mu)}={1\over\alpha_G} -{9\over 2\pi}\log({\mu\over
M_{st}}) +{.41\over 4\pi}$$
Solving the first two equations in (3.9),  one finds the scale at which the
$SU(2)$ and first $U(1)$ meet;
 $$ {M_u\over M_{st}}=EXP\left[
{(\Delta_{1a}-\Delta_2)\over2(b_{1a}-b_2)}\right] = 1.0567 \eqno(3.10)$$
Similarly for $SU(2)$ and the other four $U(1)\rm 's$, we get ${M_U\over
M_{st}}=1.0568$.
The two independent $U(1)\rm 's$  meet at ${M_U\over M_{st}}=1.057$.

For example 2, which has the same particle content as example 1,
we get slightly different results. The $\beta$-functions are
$b_a=(-6,27/4,9,35/4,35/4,35/4)$, and  the thresholds  are $\Delta_a=(-1.25;
0.015, 0.408, 0.319, 0.319, 0.319)$. Comparing the results for the two models,
we note that the $SU(2)$ $\beta$-functions and thresholds are the same, but the
$U(1)$ $\beta$-functions and thresholds are different.
This is not a surprising result. One  would expect different values for the
$U(1)$ $\beta$-functions and thresholds because the
charges are different. Here again, we see that the threshold corrections lead
to a small increase in the unification scale, which is defined
as the energy at which the couplings intersect. Table 1 shows the energy
($M_u/M_{str}$) at which the couplings intersect.
$$\matrix{-&SU(2) & 1st\, U(1) & 2nd\, U(1)  & 3rd,4th,5th\, U(1)\cr
SU(2)& -&  1.051 & 1.057 & 1.055\cr
1st\, U(1) & 1.051 & -& 1.091 & 1.078\cr
2nd\, U(1) & 1.057 & 1.091& -& 1.19\cr}$$
\centerline{Table 1- Change in unification ($M_u/M_{str}$)}
\bigskip
Although both models have the same particle content, they have different
boundary conditions on the internal fermions. For example, the second model
contains
$\nu$ values like 1/8 etc..
They also have different $\beta$ functions and different threshold corrections.
This shows that it is possible to obtain two N=1 supersymmetric models with the
same massless spectrum but  different boundary conditions on the internal
fermions. This feature allows one to modify the coupling without changing the
particle spectrum.
\vskip20pt
\centerline{\bf 4. Enhancement of gauge symmetry and mixing angle}

   In the standard model, the Weinberg angle arises from a rotation of the
third component of SU(2) with the U(1) generator. An
analogous scenario can be found with the group $SU(3)\times U(1)\times U(1)$.
In this
case one  rotates the two U(1) charges. In the model discussed here, we also
get an enhancement of the $SU(3)\times U(1)\times U(1)$ gauge symmetry to
$SU(3)\times SU(2)\times U(1)$. This arises from the
twisted fermions in the theory. Where we would normally find  ten gauge bosons
in an $SU(3)\times U(1)\times U(1)$ theory, we find twelve; two additional
spin($\pm 1$) particles
coming from the twisted sectors. A rotation of the $U(1)$ generators will give
these states quantum numbers similar to the W bosons of the standard model.
\vskip1pt
    First, we give the  Kac-Moody algebra for $SU(3)\times U(1)\times U(1)$
[24].

$$J^a(z)=\bar J^a(z)+q^a(z)$$
$$\bar J^a(z)=-{i\over 2}f_{abc}b^b(z)b^c(z)\quad ;\quad
q^q(z)=i\lambda^a_{ij}:f^i(z)\tilde f^j(z):\eqno(4.1)$$
$$J^{11}(z)={1\over \sqrt{3}}[ :f^i(z)\tilde f^i(z) : + 3\nu_2]\quad ;\quad
J^{12}(z)=[:f^1(z)\tilde f^1(z) : + \nu_1]$$
where the structure constants of SU(3) are normalized as
$f_{abc}f_{abe}=\delta_{ce}\bar C_\psi$,\vskip1pt
\noindent $3\le a,b,c \le 10$, and $2\le i,j\le 4$.
\noindent$\lambda^a_{ij}$ is an antihermitian representation of SU(3):
$$\lambda^{a*}_{ij}=-\lambda^a_{ji}\quad;\quad
[\lambda^a,\lambda^b]=f_{abc}\lambda^c\quad ;\quad
Tr\lambda^c\lambda^e=-\delta_{ce}{C^{(q)}_\psi\over 2}\eqno(4.2)$$
where $$C_\psi\equiv \bar C_\psi +C^{(q)}_\psi\quad ;\quad {C_\psi \over \bar
C_\psi}={4\over 3}\,.$$
\noindent  These equations fix the normalization of the $U(1)$ charges.

\vskip10pt

\vskip2pt
We now present a N=4 $SU(3)\times U(1)\times U(1)$ model. The number of
sectors depends on the actual values for $\nu_1$ and $\nu_2$. We first
consider a model with undetermined values of $\nu$ and derive the values that
give a unitary theory and enhance the symmetry. The generators are described by
three 16-dimensional vectors  given in (4.3) and illustrated in Fig. 3.

$$\eqalign{\rho_{b_0}& =((1)^{12};(1)^2(1)^6;(1)^8;(1)^{12})\cr
\rho_{b_1}& =((0)^{12};(0)^2(0)^6;(1)^8;(0)^{12})\cr
\rho_{b_2}&=((0)^{12};(2\nu_1)^2(2\nu_2)^6;(0)^8(0)^{12})\cr}\eqno(4.3)$$

Any superstsring theory must be unitary. It has been shown [18] that the
necessary requirements for unitarity and modular invariance are:
$${\cal N}_i\rho_{b_i}\cdot \rho_{b_i}\equiv 0\; {\rm mod}\; 8$$
for ${\cal N}_i$ even:\hskip310pt (4.4)
\vskip1pt
and $${\cal N}_{ij}\rho_{b_i}\cdot \rho_{b_j}\equiv 0\; {\rm mod}\;4$$
where ${\cal N}_{ij}$ is the least common multiple of ${\cal N}_i$ and ${\cal
N}_j$.
\vskip1pt
The additional gauge bosons come from massless states where twisted fermions
act on the vacuum. Since we want particles that are counterparts to the $W\pm$
bosons of the standard model, the massless states that will transform to these
gauge bosons should be singlets of $SU(3)$. This is accomplished by  requiring
that the massless states be excited states of the projection $\nu_1$. The mass
equation for the left movers now becomes:
$$\eqalign{\alpha' m_L^2|S>=&\,(L_0^L-1/2)|S>=0\cr 0 =&\,(\sum_{r\in Z+\lambda}
r:\tilde f_{-r}f_r: +{1\over 4}2{\nu_1}^2 +{1\over
4}6{\nu_2}^2)|0>\cr}\eqno(4.5)$$
In order for these states to survive the projections, (3.2) must also be
satisfied.
If we let $\nu_2={l\over n}$
then the conditions above,(4.4), (4.5), and (3.2) give:
$$\nu_1=1\pm \sqrt{1-3{\nu_2}^2}\eqno(4.6)$$
and the diophantine equation:
$$n^2-3l^2=m^2\eqno(4.7)$$
where n, m, l $\in Z$.
Thus we find that many different boundary conditions are allowed for this
particular set of generators. Now we will show that these boundary conditions
give rise to different mixing angles.
This is illustrated with a model with definite boundary conditions that satisfy
(4.7). I choose $\nu_1={1\over 14}$ and $\nu_2={3\over 14}$.
This model has 56 sectors and the  massless particle spectrum is:

Sector  $b_1,\phi $

        $\{\pm 2,4(\pm 3/2),6(\pm 1),4(\pm 1/2),2(0)\}$ in singlet
\vskip 1pt
        $\{\pm 1,4(\pm 1/2),6(0\}$\hskip20pt  in adjoint

\vskip5pt
Sector $b_2,b_2b_1$:
\vskip1pt
 $\{\pm 1,4(\pm1/2),6(0)\}$\hskip30pt  in $(1,{1\over
\sqrt{ 3}}9/14,-13/14)$
\vskip1pt
Sector  $b_2^{13},b_2^{13}b_1$:
\vskip1pt
  $\{\pm 1,4(\pm1/2),6(0)\}$\hskip30pt in $(1,-{1\over
\sqrt {3}}9/14,13/14)$
\vskip10pt

The last two states are the extra spin($\pm 1$) particles. We now perform
a rotation of the generators to change the quantum numbers of these states. We
  define linear orthogonal transformations of $U(1)\times U(1)$ as:
$$U(1)^\prime_y=\cos\theta\, U(1)_y +\sin\theta\, U(1)_w$$
$$T_3=U(1)^\prime_w=-\sin\theta\, U(1)_y +cos\theta\, U(1)_w\eqno(4.8)$$
and recall that in the standard model, the isospin and hypercharge are related
to the electric charge as $Q=T_3+{Y\over 2}$. In order to relate the generators
defined above in (4.8) to the generators of the standard model, we invert the
expressions in (4.8) and compare them with $Q=T_3 + {Y\over2}$ and $Q'=T_3\cot
\theta-{Y\over 2}\tan \theta$, where Q is the generator associated with $A_\mu$
and $Q\rm'$is the generator associated with $Z_\mu$.  Both sets of equations
have the same form if we let $Q=-U(1)_y/\sin\theta$, and $Y/2=-\cot\theta
U(1)'_y$.

We determine $\sin\theta$, $\cos\theta$ in (4.8) by requiring that the quantum
numbers of the massless states with spin $(\pm 1)$ have similar quantum numbers
to the gauge bosons  of the standard model. In Table 2, it is shown how the
states of this model should transform so as to have quantum numbers like the
Z,W$\pm$, and photon of the standard model.
\vfil\eject
$$\matrix{{\rm state} &   SU(3)\times U(1)_y\times U(1)_w  &
     SU(3)\times U(1)^\prime_y\times U(1)^\prime_w\cr
& & &\cr
\tilde f_{-{3\over 7}}|0>_L\otimes\epsilon \cdot b_{-{1\over2}}|0>_R&(1:{1\over
\sqrt{3}}({9\over 14}):-{13\over 14})&(1;0;-1)\cr
& & &\cr
f_{-{3\over 7}}|0>_L\otimes\epsilon \cdot b_{-{1\over2}}|0>_R &(1;-{1\over
\sqrt{3}}({9\over 14});{13\over
14})&(1;0;1)\cr& & &\cr b^{11}_{-{1\over 2}}|0>_L\otimes\epsilon \cdot
b_{-{1\over2}}|0>_R &(1;0;0) &(1;0,0)\cr & & & \cr b^{12}_{-{1\over 2}}|0>_L
\otimes\epsilon \cdot b_{-{1\over2}}|0>_R&(1;0;0) &(1;0;0)}$$
\vskip5pt
\centerline{\sl Table 2- Quantum numbers of gauge bosons }
\bigskip
For a general model with boundary conditions $\nu_1$ and $\nu_2$, there are
also states like that shown in Table 2. For the general case, the quantum
numbers can be written in terms of the boundary conditions $\nu_1$ and $\nu_2$.
For a state like the first state in table 2, the $U(1)$ quantum numbers are
$U(1)_y= \sqrt{3}\nu_2$ , and $U(1)_w=-1+\nu_1$. It is now possible to
determine the rotation required by solving
(4.8) for a state with these quantum numbers.
$$ {\textstyle{\sqrt{3} }}\nu_2\cos{\theta} +(-1+\nu_1)\sin {\theta }= 0$$
$$-{\textstyle{ \sqrt{3}}}\nu_2\sin {\theta} +(-1+\nu_1)\cos{\theta}=
-1\eqno(4.9)$$
Solving this for $\cos \theta$ and $\sin \theta$, one  obtains:
$$\cos\theta =-(-1+\nu_1)$$
$$\sin\theta =\sqrt{3}\,\nu_2\eqno(4.10)$$
$\sin\theta$ is analogous to the Weinberg angle.

 In the N=4 supersymmetric model presented above, the symmetry group of the
standard model arises by an enhancement of the gauge symmetry $SU(3)\times
U(1)\times U(1)$. This enhancement was brought about by the complex twisted
fermions of the model.
One can obtain  gauge bosons in adjoint representations by making a rotation of
the gauge generators as in (4.8). This
gives rise to a  mixing angle analogous to the Weinberg angle of the standard
model. It has been shown that this mixing angle depends on the boundary
conditions of the internal fermions as in (4.10). Once this has been determined
for a realistic  string model, the $\beta $ -function and threshold corrections
would allow one to calculate the couplings at measurable energies. Without
further restrictions, one could choose different boundary conditions to give an
acceptable mixing angle.
\vskip5pt

\centerline{ {\bf 5. Conclusion}}

This paper is concerned with the calculation of threshold corrections in
twisted string models and how they might change the unification point. For the
two
N=1 $SU(2)\times U(1)^5$ models presented, the threshold corrections lead to a
very small increase in the unification scale. In
the examples chosen here,  all moduli are fixed to be at the point where the
gauge symmetry is enlarged. So, unlike heterotic models,
there are no free parameters which can be adjusted to give large thresholds.

As a further step in making a connection with low energy phenomenology, we
showed how the complex twisted fermions
can enhance the symmetry of an N=4 $SU(3)\times U(1)\times U(1)$ model to the
gauge group $SU(3)\times SU(2)\times U(1)$.
The boundary conditions on the twisted fermions of this model give massless
spin $\pm 1$ particles that
are singlets of $SU(3)$ and  a model that is unitary and modular invariant. We
find, for this particular
model at least, that there are many possible boundary conditions that would
have the same particle spectrum. When these massless
spin$\pm 1$ particles are required to have quantum numbers like the gauge
bosons of the standard model, a mixing angle analogous to the Weinberg angle
can be identified. An expression can then be derived for the dependence of this
mixing angle on the
internal boundary conditions of the model. Once more realistic string models
are developed, the methods used here should allow one to compare the string
theory with measurable quantities. Using the $\beta$- function and string
thresholds, one could determine the gauge couplings
at low energies. If there are no additional constraints on the boundary
conditions of the internal fermions, one would have
the freedom to choose a model that best describes the low energy physics.

\bigskip
\noindent I would like to thank my advisor Louise Dolan,  and J.T. Liu for
helpful discussions.
\bigskip
\noindent This work is supported in part by the US Department of Energy under
grant
DE-FG05-35ER-40219

\centerline{   References}
\item{1.} S. Weinberg, Phys. Lett. {\bf B91} (1980) 51.
\item{2.} P.Langacker and M. Luo. Phys. Rev. {\bf D44} (1991) 817.
\item{3.} T.R.Taylor and B.Veneziano, Phys. Lett. {\bf B212} (1988) 14.
\item{4.} J.A. Minahan, Nucl. Phys. {\bf B298} (1988) 36.
\item{5.} V.S. Kaplunovsky, Nucl. Phys. {\bf B307} (1988) 145 [Erratum:{\bf
B382} (1992) 436].
\item{6.} L.J. Dixon, V.S. Kaplunovsky and J. Louis, Nucl. Phys. {\bf B355}
(1991) 649.
\item{7.} L.E. Ibanez,D. Lust and G.G.Ross,Phys. Lett. {\bf B272} (1991) 251.
\item{8.} L.F. Abbott, Nucl. Phys. {\bf B185 }(1981) 189.
\item{9.} I. Antoniadis, J. Ellis, R.Lacaze and D. Nanopoulos, Phys. Lett. {\bf
B268 }(1991) 188.
\item{10.} A. Font, L.E. Ibanez, D.Lust and F. Quevedo, Phys. Lett. {\bf B245}
(1990) 401.
\item{11.} S. Ferrara, N. Magnoli, T.R. Taylor and G. Veneziano, Phys. Lett.
{\bf B245} (1990) 409.
\item{12.} I. Antoniadis, K.S. Narain and T.R. Taylor, Phys. Lett. {\bf B267}
(1991) 37.
\item{13.} J.A. Cases and C. Munoz, Phys. Lett. {\bf B271} (1991) 85.
\item{14.} T. Banks,L. Dixon, D.Friedan, E. Martinec, Nucl. Phys.  {\bf B299}
(1988) 613.
\item{15.} P.Ginsparg, Phys. Lett. {\bf B197} (1987) 139.
\item{16.} J. Polchinski, Comm. Math Phys. {\bf 104 } (1986) 27.
\item{17.} L.Dolan, J.T. Liu, Nucl. Phys. {\bf B387} (1992) 86.
\item{18.} R. Bluhm, L. Dolan and P. Goddard, Nucl. Phys. {\bf B289} (1987)
364.
\item{19.} R. Bluhm, L. Dolan and P. Goddard, Nucl. Phys. {\bf B309} (1988)
330.
\item{20.} I.Antoniadis, C.Bachas and C.Kounnas, Nucl. Phys. {\bf B289} (1987)
87.
\item{21.} H. Kawai, D. Lewellen, J.A. Schwartz and H. Tye, Nucl. Phys. {\bf
B299} (1988) 431.
\item{22.} D. Gross, F. Wilczek, Phys. Rev. {\bf D8} (1973) 3633.
\item{23.} L.J. Dixon,V.S.Kaplunovsky and C.Vafa, Nucl.Phys. {\bf B294} (1987)
43.
\item{24.} L.Dolan in Strings '88, J.Gates and W.Seigel eds., Singapore: World
Scientific(1988).

\end